\documentstyle[12pt]{article}

\catcode`\@=11
\long\def\@makefntext#1{
\protect\noindent \hbox to 3.2pt {\hskip-.9pt
$^{{\ninerm\@thefnmark}}$\hfil}#1\hfill}                

 \def\@makefnmark{\hbox to 0pt{$^{\@thefnmark}$\hss}}  

\def\ps@myheadings{\let\@mkboth\@gobbletwo
\def\@oddhead{\hbox{}
\rightmark\hfil\ninerm\thepage}
\def\@oddfoot{}\def\@evenhead{\ninerm\thepage\hfil
\leftmark\hbox{}}\def\@evenfoot{}
\def\sectionmark##1{}\def\subsectionmark##1{}}

\newcounter{sectionc}\newcounter{subsectionc}\newcounter{subsubsectionc}
\renewcommand{\section}[1] {\vspace{0.6cm}\addtocounter{sectionc}{1}
\setcounter{subsectionc}{0}\setcounter{subsubsectionc}{0}\noindent
	{\bf\thesectionc. #1}\par\vspace{0.4cm}}
\renewcommand{\subsection}[1] {\vspace{0.6cm}\addtocounter{subsectionc}{1}
	\setcounter{subsubsectionc}{0}\noindent
	{\it\thesectionc.\thesubsectionc. #1}\par\vspace{0.4cm}}
\renewcommand{\subsubsection}[1]
{\vspace{0.6cm}\addtocounter{subsubsectionc}{1}
	\noindent {\rm\thesectionc.\thesubsectionc.\thesubsubsectionc.
	#1}\par\vspace{0.4cm}}

\newcounter{appendixc}
\newcounter{subappendixc}[appendixc]
\newcounter{subsubappendixc}[subappendixc]

\renewcommand{\appendix}[1] {\vspace{0.6cm}
	\refstepcounter{appendixc}
	\setcounter{figure}{0}
	\setcounter{table}{0}
	\setcounter{equation}{0}
	\renewcommand{\thefigure}{\Alph{appendixc}.\arabic{figure}}
	\renewcommand{\thetable}{\Alph{appendixc}.\arabic{table}}
	\renewcommand{\theappendixc}{\Alph{appendixc}}
	\renewcommand{\theequation}{\Alph{appendixc}.\arabic{equation}}
	\noindent{\bf Appendix \theappendixc #1}\par\vspace{0.4cm}}

\def\abstracts#1{{
	\centering{\begin{minipage}{30pc}\tenrm\baselineskip=12pt\noindent
	\centerline{\tenrm ABSTRACT}\vspace{0.3cm}
	\parindent=0pt #1
	\end{minipage}}\par}}

\renewenvironment{thebibliography}[1]
	{\begin{list}{\arabic{enumi}.}
	{\usecounter{enumi}\setlength{\parsep}{0pt}
\setlength{\leftmargin 1.25cm}{\rightmargin 0pt}
	 \setlength{\itemsep}{0pt} \settowidth
	{\labelwidth}{#1.}\sloppy}}{\end{list}}

\newcounter{itemlistc}
\newcounter{romanlistc}
\newcounter{alphlistc}
\newcounter{arabiclistc}

\newcommand{\fcaption}[1]{
	\refstepcounter{figure}
	\setbox\@tempboxa = \hbox{\tenrm Fig.~\thefigure. #1}
	\ifdim \wd\@tempboxa > 6in
	   {\begin{center}
	\parbox{6in}{\tenrm\baselineskip=12pt Fig.~\thefigure. #1}
	    \end{center}}
	\else
	     {\begin{center}
	     {\tenrm Fig.~\thefigure. #1}
	      \end{center}}
	\fi}

\newcommand{\tcaption}[1]{
	\refstepcounter{table}
	\setbox\@tempboxa = \hbox{\tenrm Table~\thetable. #1}
	\ifdim \wd\@tempboxa > 6in
	   {\begin{center}
	\parbox{6in}{\tenrm\baselineskip=12pt Table~\thetable. #1}
	    \end{center}}
	\else
	     {\begin{center}
	     {\tenrm Table~\thetable. #1}
	      \end{center}}
	\fi}

\def\@citex[#1]#2{\if@filesw\immediate\write\@auxout
	{\string\citation{#2}}\fi
\def\@citea{}\@cite{\@for\@citeb:=#2\do
	{\@citea\def\@citea{,}\@ifundefined
	{b@\@citeb}{{\bf ?}\@warning
	{Citation `\@citeb' on page \thepage \space undefined}}
	{\csname b@\@citeb\endcsname}}}{#1}}

\newif\if@cghi
\def\cite{\@cghitrue\@ifnextchar [{\@tempswatrue
	\@citex}{\@tempswafalse\@citex[]}}
\def\citelow{\@cghifalse\@ifnextchar [{\@tempswatrue
	\@citex}{\@tempswafalse\@citex[]}}
\def\@cite#1#2{{$\null^{#1}$\if@tempswa\typeout
	{IJCGA warning: optional citation argument
	ignored: `#2'} \fi}}

\def\fnt#1#2{\footnotetext{\kern-.3em
	{$^{\mbox{\sevenrm #1}}$}{#2}}}

 1
 1
 1

\font\tenbf=cmbx10
\font\tenrm=cmr10
\font\tenit=cmti10

\font\ninerm=cmr9

\begin{document}

\centerline{\tenbf THE FERMION MASS PROBLEM
\footnote{To be published in the Proceedings of the 7th Adriatic Meeting
on Particle Physics: Perspectives in Particle Physics '94, Islands of Brijuni,
Croatia, 13 - 20 September 1994.}}
\vspace{0.2cm}
\centerline{\tenrm C. D. FROGGATT}
\baselineskip=13pt
\centerline{\tenit Department of Physics and Astronomy, University of Glasgow}
\baselineskip=12pt
\centerline{\tenit Glasgow, G12 8QQ, Scotland, U.K.}
\vspace{0.4cm}
\abstracts{Different approaches to the quark-lepton mass problem are reviewed.
The infrared quasifixed point predictions for the top quark mass
are discussed for the Standard Model and its minimal supersymmetric extension,
with particular reference to the large $\tan\beta$ scenario and Yukawa
unification. Mass matrix ans\"{a}tze with texture zeros at the unification
scale are also considered. It is argued that the hierarchy of fermion masses
and mixing angles requires the existence of an approximately conserved chiral
flavour symmetry beyond the Standard Model.}
\vspace{0.8cm}
\rm\baselineskip=14pt
\section{Introduction}
One of the most important unresolved problems of particle physics is the
understanding of flavour and the fermion mass spectrum. The observed values
of the quark and lepton masses and the quark mixing angles provide our
main experimental clues to the underlying flavour dynamics contained in the
physics beyond the Standard Model (SM). The most striking qualitative features
of the spectroscopy of quarks and charged leptons are:

\begin{enumerate}
\item
The fermion mass hierarchy: the large mass ratios of order 60 between fermions
of a given electric charge, i.~e.\ of the same family.
\item
The fermion generation structure: the similarity between the mass spectra of
the three families of quarks and charged leptons.
\item
The quark mixing hierarchy: the smallness of the off-diagonal elements of the
quark weak coupling matrix $V_{CKM}$.
\end{enumerate}
Overall the charged fermion masses range over five orders of magnitude, from
\mbox{1/2 Mev} for the electron to over 100 GeV for the top quark.

A three generation structure is clearly indicated, consisting of
\mbox{(u,d,e,$\nu_e$)}, \mbox{(c,s,$\mu,\nu_\mu$)} and
\mbox{(t,b,$\tau,\nu_\tau$)} respectively. As is well known, each generation
forms an anomaly free representation of the SM gauge group (SMG). The LEP
measurements of the Z width show that there are just three neutrinos
with masses less than $M_{Z}/2$ or more
\hspace{5pt} precisely \cite{pdg}
\begin{equation}
N_\nu = 2.985 \pm 0.023 \pm 0.004
\end{equation}
We conclude that there are three generations of quarks and leptons, unless
there
exists (i) a heavy neutrino at the electroweak scale \cite{hillpas} or
(ii) a fourth generation of quarks without leptons, but having the SM gauge
anomalies cancelled against those of a
generation of 'techniquarks', associated with an extra non-abelian gauge group
extending the SMG \cite{douglas}.

Neutrino masses, if non-zero, would seem to have a different origin to those of
the quarks and charged leptons. In the SM there are no right-handed weak
isosinglet neutrino states $\nu_R$ and the Higgs mechanism cannot generate a
neutrino mass term. In extensions of the SM it is possible to generate Majorana
mass terms connecting the left-handed weak isodoublet neutrinos of the SM with
the corresponding set of right-handed weak isodoublet anti-neutrinos. These
Majorana mass terms break weak isospin by one unit ($\Delta t = 1$) as well
as lepton flavour conservation. Such a $\Delta t = 1$ mass term can be
generated by: (i) the exchange of the the usual Higgs tadpole
$\langle\phi_{WS}\rangle$ twice, via a superheavy lepton $L^0$ intermediate
state having the same gauge quantum numbers as $\nu_R$ (i.~e.\ neutral)
under the SM \cite{fn2,seesaw}; or (ii) the exchange of a single weak
isotriplet
Higgs tadpole \cite{gelmini}. Method (i) has become known as the see-saw
mechanism, since it generates a neutrino mass scale of
\mbox{$\langle\phi_{WS}\rangle^2/M_{L^0}$}, suppressed by a factor of
\mbox{$\langle\phi_{WS}\rangle/M_{L^0}$} relative to the natural charged
fermion mass scale of \mbox{$\langle\phi_{WS}\rangle = 174$ Gev.} More details
about neutrino masses will be found in other contributions to this meeting
\cite{halzen,smirnov}.

Here we are really concerned with the charged fermion mass problem and the
three main approaches to it:
\begin{enumerate}
\item
Attempts to derive a fermion mass or mass relation exactly
from some dynamical or theoretical principle.
\item
Searches for relationships between mass and mixing angle
parameters using symmetries and/or ans\"{a}tze to make detailed
fits to the data.
\item Attempts to naturally explain all the qualitative features of
the fermion spectrum, fitting all the data within factors of order unity.
\end{enumerate}

We shall illustrate these approaches by reviewing some recent developments in
models of the quark and lepton mass matrices. An example of a mass relation
following from an a priori theoretical principle is Veltman's condition
\cite{veltman}
\begin{equation}
\sum_{leptons}m_l^2 + \sum_{quarks}m_q^2 = \frac{3}{2} M_W^2 + \frac{3}{4}
M_Z^2
+ \frac{3}{4} M_H^2
\end{equation}
for the cancellation of quadratic divergences to one loop in the SM. In the
next
section we will consider predictions of the top quark mass based on the strong
coupling dynamics of a renormalisation group infrared fixed point. The Fritzsch
ansatz and its generalisation to mass matrix ans\"{a}tze with texture zeros, in
the context of supersymmetric grand unified (SUSY-GUT) models, will be
considered
as examples of the second approach. Finally we will turn to mass protection, by
chiral flavour charges beyond the SM, for a natural explanation of the fermion
mass hierarchy. We will consider examples of this last approach based on the
minimal supersymmetric Standard Model (MSSM), SUSY-GUTS and antigrand
unification.

\section{Renormalisation Group Fixed Points and the Top Quark Mass}
The idea that properties of the observed fermion mass spectrum could be
explained in
terms of an infrared fixed point of the renormalisation group equations (RGE)
for the Yukawa coupling constants was first considered \cite{fn1} some time
ago.
It was pointed out that the three generation fermion
mass hierarchy does not develop naturally out of the general structure of
the RGE. However it was soon realised \cite{pendleton} that
the top quark mass might correspond to a fixed point value of the SM RGE,
predicting approximately \mbox{$m_{t} \simeq 100$ Gev}
\cite{pendleton,marciano}.
In practice one finds that such an infrared fixed point behaviour of the
running top quark Yukawa coupling constant $g_{t}(\mu)$ does not generically
set in until \mbox{$\mu < 1$ Gev}, where the QCD coupling constant $g_{3}(\mu)$
varies rapidly. The scale relevant for the physical top quark mass prediction
is of course $\mu = m_{t}$; at this scale $g_{3}(\mu)$ is slowly varying and
there is an effective infrared stable quasifixed point (which would be an exact
fixed point if $g_{3}(\mu)$ were constant) behaviour giving a
running top quark mass prediction \mbox{$m_{t}(\mu = m_{t}) \simeq 225$ Gev}
\cite{hill}.

More precisely the SM quasifixed point prediction for the top quark mass
requires the following assumptions:
\begin{enumerate}
\item
The desert hypothesis of no new interactions beyond those of the SM up to some
high energy scale \mbox{$\mu = M_{X} \simeq 10^{15} - 10^{19}$ Gev}, e.~g.\ the
grand unification scale or the Planck scale.
\item
The SM coupling constants remain positive and finite in the desert, such that
perturbation theory and the RGE can be applied up to \mbox{$\mu = M_{X}$}.
\item
The top quark Yukawa coupling constant is large at \mbox{$\mu = M_{X}$}:
\begin{equation}
1 \leq g_{t}(M_{X}) \leq \sqrt{4\pi}
\end{equation}
so that it enters the domain of attraction of the infrared quasifixed point.
\end{enumerate}
The nonlinearity of the RGE then strongly focuses $g_{t}(\mu)$ at the
electroweak scale to its quasifixed point value. The RGE for the Higgs
self-coupling $\lambda(\mu)$ similarly focuses $\lambda(\mu)$ towards a
quasifixed point value, leading to the SM fixed point predictions for the
running top quark and Higgs masses:
\begin{equation}
m_{t} \simeq 225\ \mbox{Gev} \quad m_{H} \simeq 250\ \mbox{Gev}
\end{equation}
Unfortunately  the
LEP results \cite{LEP} and the CDF measurement \cite{CDF}, which require a
running top mass \mbox{$m_{t} \simeq 165 \pm 15$ Gev} are inconsistent with
this fixed point prediction for the top quark mass. Note that the running
quark mass $m_{q}$ is related to the  physical or pole quark mass $M_{q}$,
defined as the location of the pole in the quark propagator, by
\begin{equation}
M_{q} = m_{t}(M_{q})(1 + 4\alpha_{3}(M_{q})/3\pi)
\end{equation}
at the one loop QCD level.

There are two interesting modifications to the fixed point top mass
prediction in the minimal supersymmetric Standard Model (MSSM) with
supersymmetry breaking at the electroweak scale or Tev scale:
\begin{itemize}
\item
The introduction of the supersymmetric partners of the SM particles in the
RGE for the Yukawa and gauge coupling constants leads to a 15\% reduction in
the fixed point value of $g_{t}(m_{t})$ \cite{bagger,dimhallrab}.
\item
There are two Higgs doublets in the MSSM and the
ratio of Higgs vacuum values, $\tan \beta = v_{2}/v_{1}$, is a free parameter;
the top quark couples to $v_{2}$ and so $m_{t}$ is proportional to
\mbox{$v_{2} = (174\ \mbox{Gev})\sin\beta$}.
\end{itemize}
The MSSM fixed point prediction for the running top quark mass is
\cite{barger}:
\begin{equation}
m_{t}(m_{t}) \simeq (190\ \mbox{Gev})\sin\beta
\end{equation}
which is remarkably close to the LEP and CDF results for
\mbox{$\tan\beta > 1$}
This quasifixed point value is of course also the upper bound on the top mass
in the MSSM, assuming perturbation theory is valid in the desert up to the
SUSY-GUT scale. It then follows  that the experimental
evidence for a large top mass requires
\mbox{$\tan\beta > 1$}. We note that the minimal SU(5) SUSY-GUT symmetry
relation between the bottom quark and tau lepton Yukawa coupling constants,
\mbox{$g_{b}(M_{X}) = g_{\tau}(M_{X})$}, is also only satisfied
phenomenologically
if the top quark Yukawa coupling is close to its infrared quasifixed point
value,
so that it contributes significantly to the running of \mbox{$g_{b}(\mu)$} and
reduces the predicted value of $m_{b}(m_{b})$. In the SM the contribution
of the top quark Yukawa coupling has the opposite sign and the SU(5) GUT
prediction for $m_{b}(m_{b})$ fails, as it is then phenomenologically too
large.

For large $\tan\beta$ it is possible to have a bottom quark Yukawa coupling
satisfying \mbox{$g_{b}(M_{X}) \geq 1$} which then approaches an infrared
quasifixed point and is no longer negligible in the RGE for $g_{t}(\mu)$.
Indeed with
\begin{equation}
\tan\beta \simeq m_{t}(m_{t})/m_{b}(m_{t}) \simeq 60
\end{equation}
we can trade the mystery of the top to bottom quark mass ratio
for that of a
hierarchy of vacuum expectation values, \mbox{$v_{2}/v_{1} \simeq
m_{t}(m_{t})/m_{b}(m_{t})$},
 and have all the third generation Yukawa coupling constants large:
\begin{equation}
g_{t}(M_{X}) \geq 1 \quad g_{b}(M_{X}) \geq 1 \quad g_{\tau}(M_{X}) \geq 1
\label{tbtaufp}
\end{equation}
Then $m_{t}$, $m_{b}$ and \mbox{$R = m_{b}/m_{\tau}$} all approach infrared
quasifixed point
values compatible with experiment \cite{fkm}. This large $\tan\beta$
scenario is consistent with the idea of Yukawa unification \cite{anan}:
\begin{equation}
g_{t}(M_{X}) = g_{b}(M_{X}) = g_{\tau}(M_{X}) = g_{G}
\label{yukun}
\end{equation}
as occurs in the SO(10) SUSY-GUT model with
the two MSSM Higgs doublets in a single {\bf 10} irreducible representation
and $g_{G} \ge 1$ ensures fixed point behaviour.
However it should be noted that the equality in Eq.~(\ref{yukun}) is not
necessary. For example in SU(5) finite unified theories \cite{zoupanos}
the Yukawa couplings are related to the SUSY-GUT coupling constant and
satisfy \mbox{$g_{t}^{2}(M_{X}) = 4g_{b}^{2}(M_{X})/3$ = $\cal O$(1)}, giving
the same fixed point predictions.
In fact one does not need a symmetry assumption at all, since the weaker
assumption of large third generation Yukawa couplings, Eq.~(\ref{tbtaufp}),
is sufficient for the fixed point dynamics to predict \cite{fkm}
the running masses
$m_{t} \simeq 180 \ \mbox{Gev}$, $m_{b} \simeq 4.1 \ \mbox{Gev}$ and
$m_{\tau} \simeq 1.8 \ \mbox{Gev}$ in the large $\tan\beta$ scenario. Also
the lightest Higgs particle mass is predicted to be $m_{h^0} \simeq 120 \
\mbox{Gev}$ (for a top squark mass of order  \mbox{1 Tev}).

The origin of the large value of \mbox{$\tan\beta$} is of course a puzzle,
which
must be solved before the large \mbox{$\tan\beta$} scenario can be said to
explain the large \mbox{$m_{t}/m_{b}$} ratio. It is possible to introduce
approximate symmetries \cite{anderson,hall} of the Higgs potential which
ensure a hierarchy of vacuum expectation values - a Peccei-Quinn symmetry and
a continuous $\cal R$ symmetry have been used. However these symmetries then
result in a light chargino \cite{nelson}, in conflict with the LEP
lower bound of order \mbox{45 Gev} on the chargino mass,
unless the SUSY breaking scale $M_{SUSY}$ is fine-tuned to be much
larger than the electroweak scale: \mbox{$M_{SUSY}^2 \geq \tan\beta M_{Z}^2$}.
The Peccei-Quinn and $\cal R$ symmetries require a hierarchical SUSY
spectrum with the squark and slepton masses much larger than the gaugino,
Higgsino and Z masses. In particular they are inconsistent with the popular
scenario of universal soft SUSY breaking mass parameters at the unification
scale and radiative electroweak symmetry breaking \cite{carena}.

Also, in the large $\tan\beta$ scenario, SUSY radiative corrections to $m_{b}$
are generically large: the bottom quark mass gets a contribution proportional
to $v_{2}$ from some one-loop
diagrams with internal superpartners, such as top squark-charged Higgsino
exchange , whereas its tree level mass is proportional to
$v_{1} = v_{2}/\tan\beta$. Consequently these loop diagrams give a
fractional correction \mbox{$\delta m_{b}/m_{b}$} to the bottom quark mass
proportional to $\tan\beta$ and generically of order unity
\cite{hall,carena}. The presence of
the above-mentioned Peccei-Quinn and $\cal R$ symmetries and the associated
hierarchical SUSY spectrum (with the squarks much heavier than
the gauginos and Higgsinos) would protect $m_{b}$ from large radiative
corrections, by providing a suppression factor in the loop diagrams and
giving \mbox{$\delta m_{b}/m_{b} \ll 1$}. The hierarchical
superpartner mass spectrum would also suppress a similar
$\cal O(\tan\beta)$ enhancement of the rare $b \rightarrow s \gamma$ decay
amplitude, which would otherwise be in conflict with the CLEO data \cite{cleo}.
However, in the absence of
experimental information on the superpartner spectrum, the predictions of the
third generation quark-lepton masses in the large $\tan\beta$ scenario
must, unfortunately, be considered unreliable.

\vspace{-1pt}
\section{Mass Matrix Ans\"{a}tze and Texture Zeros}
\vspace{-1pt}
The motivation for considering mass matrix ans\"{a}tze is to obtain
testable relationships between fermion masses and mixing angles, thereby
reducing the number of free parameters in the SM and providing a hint to
the physics beyond the SM.
The best known ansatz for the quark mass matrices
is due to Fritzsch \cite{fritzsch}:

\begin{equation}
M_U =\pmatrix{0  		& C   		& 0\cr
		      C  		& 0   		& B\cr
		      0  		& B   		& A\cr}
\qquad
M_D =\pmatrix{0  		& C^\prime  & 0\cr
		      C^\prime 	& 0   		& B^\prime\cr
		      0  		& B^\prime  & A^\prime\cr}
\end{equation}
It contains 6 complex parameters A,B,C,$A^\prime$,$B^\prime$ and
$C^\prime$. Four of the phases can be rotated away by redefining the phases of
the quark fields, leaving just 8 real parameters (the magnitudes of
A,B,C,$A^\prime$,$B^\prime$ and $C^\prime$ and two phases
$\phi_{1}$ and $\phi_{2}$) to reproduce 6 quark masses and
4 angles parameterising $V_{CKM}$. There are thus two relationships predicted
by the Fritzsch ansatz.
It is necessary to {\em assume}:
\begin{equation}
|A| \gg |B| \gg |C|, \qquad |A^\prime| \gg |B^\prime| \gg |C^\prime|
\end{equation}
in order to obtain a good fermion mass hierarchy.

The first prediction is a generalised version of the relation
$\theta_c\simeq\sqrt{\frac{m_d}{m_s}}$ for the Cabibbo angle,
which originally motivated the ansatz:
\begin{equation}
|V_{us}| \simeq
\left| \sqrt{\frac{m_{d}}{m_{s}}} -
e^{-i\phi_{1}}\sqrt{\frac{m_{u}}{m_{c}}} \right|
\label{fritzsch1}
\end{equation}
and is well satisfied experimentally. However the second relationship:
\begin{equation}
|V_{cb}| \simeq
\left| \sqrt{\frac{m_{s}}{m_{b}}} -
e^{-i\phi_{2}}\sqrt{\frac{m_{c}}{m_{t}}} \right|
\label{fritzsch2}
\end{equation}
cannot be satisfied with a heavy top quark. Using
\mbox{$\sqrt{\frac{m_s}{m_b}} \simeq 0.18$} and \mbox{$|V_{cb}| \leq 0.055$},
an
upper limit of \mbox{$m_{t} < 100$ Gev} is obtained \cite{gilman}. The limit
is valid in the SM whether the ansatz is applied at the electroweak scale or
at the GUT scale. This is also true in the MSSM.
So, using the standard quark masses \cite{pdg},
the Fritzsch ansatz is excluded by the data.

Recently ans\"{a}tze incorporating relationships between the fermion mass
parameters at the grand unified or the Planck scale have been studied.
We have already mentioned the best known result: the simple SU(5) relation
$m_{b}(M_{X}) = m_{\tau}(M_X)$ which is satisfied in SUSY-GUTs provided the
top quark mass is near to its quasifixed point value \cite{dimhallrab,arason}.
However the corresponding relations for the first two generations are not
satisfied, as they predict for example
\begin{equation}
m_{d}/m_{s} = m_{e}/m_{\mu}
\label{eq:massratio}
\end{equation}
which fails phenomenologically by an order of magnitude. This led Georgi and
Jarlskog \cite{georgijarlskog,harvey} to postulate the mass relations
$m_{b}(M_X) = m_{\tau}(M_{X})$, $m_{s}(M_X) = m_{\mu}(M_{X})$/3 and
$m_{d}(M_X) = 3m_{e}(M_{X})$ at the GUT scale.
Dimopoulos, Hall and Raby \cite{dimhallrab} revived these relations
in the context of an SO(10) SUSY-GUT, combining the Fritzsch form for the up
quark mass matrix \mbox{$M_{U} = Y_{u}v_{2}$} with the Georgi-Jarlskog form
for the down quark and charged lepton mass matrices
\mbox{$M_{D} = Y_{d}v_{1}$} and \mbox{$M_{L} = Y_{l}v_{1}$}:

\begin{equation}
Y_u =\pmatrix{0  		 & C   				& 0\cr
		      C  		 & 0   				& B\cr
		      0  		 & B   				& A\cr}
\quad
Y_d =\pmatrix{0  		 & Fe^{i\phi}		& 0\cr
		      F^{-i\phi} & E   				& 0\cr
		      0  		 & 0				& D\cr}
\quad
Y_l =\pmatrix{0  		& F   				& 0\cr
		      F  		& -3E  				& 0\cr
		      0  		& 0   				& D\cr}
\end{equation}
The  phase freedom in the definition of the fermion fields has been
used to make the parameters A, B, C, D, E and F real and we have again to
assume:
\begin{equation}
|A| \gg |B| \gg |C|, \qquad |D| \gg |E| \gg |F|
\label{eq:hierarchy}
\end{equation}
Thus there are 7 free parameters in the Yukawa coupling ansatz and
$\tan\beta$ available to fit 13 observables. Using the RGE from the
SUSY-GUT scale to the electroweak scale, this ansatz gives 5 predictions
which are, within errors, in agreement with data for
\mbox{$1 < \tan\beta < 60$} \cite{dimhallrab,bargander}. The failed simple
SU(5) prediction Eq.~(\ref{eq:massratio}) is replaced by the successful mass
ratio
prediction
\begin{equation}
(m_{d}/m_{s}) (1 - m_{d}/m_{s})^{-2} =
9 (m_{e}/m_{\mu}) (1 - m_{e}/m_{\mu})^{-2}
\end{equation}
Since the down quark matrix $Y_{d}$ is diagonal in the two heaviest
generations,
one of the SUSY-GUT scale predictions is \cite{harvey}
\mbox{$V_{cb} \simeq \sqrt{\frac{m_{c}}{m_{t}}}$}.
Fits give $m_{t}$ close to its fixed point and the large top
Yukawa coupling causes  $V_{cb}$ to run
between the GUT and electroweak scales to a somewhat lower value. Nonetheless
the fits still tend to make $V_{cb}$ too large.
A fit satisfying Yukawa unification is obtained by setting
\mbox{$A = D$} and \mbox{$\tan\beta \simeq 60$}. It is of course subject to
uncertainties due to the possibly large SUSY radiative corrections to $m_{b}$
mentioned in the previous section.

\begin{table}[h,t,b]
\centering
\tcaption{Approximate forms for the symmetric textures. The
parameter $\lambda \simeq 0.2$ is the CKM matrix element $V_{us}$ used in the
Wolfenstein parameterisation of $V_{CKM}$.}
\vspace{0.2cm}
\begin{tabular}{|c|c|c|} \hline
 & U & D \\ \hline
1 & $\left(
\begin{array}{ccc}

0 & \sqrt{2}\lambda^6 & 0 \\
\sqrt{2}\lambda^6 & \lambda^4 & 0 \\
0 & 0 & 1
\end{array}
\right)$ & $\left(
\begin{array}{ccc}

0 & 2\lambda^4 & 0 \\
2\lambda^4 & 2\lambda^3 & 4\lambda^3 \\
0 & 4\lambda^3 & 1
\end{array}
\right)$
\\ \hline

2 & $\left(
\begin{array}{ccc}

0 & \lambda^6 & 0 \\
\lambda^6 & 0 & \lambda^2 \\
0 & \lambda^2 & 1
\end{array}
\right)$ & $\left(
\begin{array}{ccc}

0 & 2\lambda^4 & 0 \\
2\lambda^4 & 2\lambda^3 & 2\lambda^3 \\
0 & 2\lambda^3 & 1
\end{array}
\right)$
\\ \hline

3 & $\left(
\begin{array}{ccc}

0 & 0 & \sqrt{2}\lambda^4 \\
0 & \lambda^4 & 0 \\
\sqrt{2}\lambda^4 & 0 & 1
\end{array}
\right)$ & $\left(
\begin{array}{ccc}

0 & 2\lambda^4 & 0 \\
2\lambda^4 & 2\lambda^3 & 4\lambda^3 \\
0 & 4\lambda^3 & 1
\end{array}
\right)$
\\ \hline

4 & $\left(
\begin{array}{ccc}

0 & \sqrt{2}\lambda^6 & 0 \\
\sqrt{2}\lambda^6 & \sqrt{3}\lambda^4 & \lambda^2 \\
0 & \lambda^2 & 1
\end{array}
\right)$ & $\left(
\begin{array}{ccc}

0 & 2\lambda^4 & 0 \\
2\lambda^4 & 2\lambda^3 & 0 \\
0 & 0 & 1
\end{array}
\right)$
\\ \hline

5 & $\left(
\begin{array}{ccc}

0 & 0 & \lambda^4 \\
0 & \sqrt{2}\lambda^4 & \frac{\lambda^2}{\sqrt{2}} \\
\lambda^4 & \frac{\lambda^2}{\sqrt{2}} & 1
\end{array}
\right)$ & $\left(
\begin{array}{ccc}

0 & 2\lambda^4 & 0 \\
2\lambda^4 & 2\lambda^3 & 0 \\
0 & 0 & 1
\end{array}
\right)$
\\ \hline
\end{tabular}
\label{table:ross}
\end{table}

The predictions arise due to the reduction in the number of free parameters,
obtained by requiring the presence of zeros and symmetries between mass
matrix elements. A systematic analysis \cite{texture}
of {\em symmetric} quark mass matrices
with 5 or 6 ``texture'' zeros at the SUSY-GUT scale has recently been made.
There are just 6 possible forms of symmetric mass matrix with an hierarchy
of three non-zero eigenvalues and three texture zeros. These are:
\begin{equation}
\pmatrix{a_1  		 & 0   				& 0  \cr
		 0  		 & b_1   			& 0  \cr
		 0  		 & 0   				& c_1\cr}
\qquad
\pmatrix{0  		 & a_2		& 0  \cr
		 a_2         & b_2   	& 0  \cr
		 0       	 & 0		& c_2\cr}
\qquad
\pmatrix{a_3  		& 0   		& 0  \cr
		 0      	& 0  		& b_3\cr
		 0   		& b_3   	& c_3\cr}
\end{equation}
and
\begin{equation}
\pmatrix{0  		 & 0   		& a_4\cr
		 0  		 & b_4   	& 0  \cr
		 a_4  		 & 0   		& c_4\cr}
\qquad
\pmatrix{0  		 & a_5		& 0  \cr
		 a_5         & 0   		& b_5\cr
		 0       	 & b_5		& c_5\cr}
\qquad
\pmatrix{0  		& a_6   	& b_6\cr
		 a_6      	& 0  		& 0  \cr
		 b_6   		& 0     	& c_6\cr}
\end{equation}
Comparison with the measured values of quark masses and mixing angles
yields \cite{texture} another
5 quark mass matrix ans\"{a}tze consistent with experiment. The hierarchical
structure of the parameters in the ans\"{a}tze (cf. Eq.~(\ref{eq:hierarchy}))
suggests a parameterisation of the form \cite{texture}  shown in
Table~\ref{table:ross}, analogous to that of Wolfenstein \cite{wolfenstein}
for the quark mixing matrix.
It is natural to interpret $\lambda$ as a symmetry
breaking parameter for some approximate symmetry beyond those of
the Standard Model Group (SMG). The nature of
this symmetry is discussed in the next section.

The neutrino Majorana mass matrices generated by the see-saw mechanism in
many extensions of the SM naturally have the above type of symmetric texture.
Due to the hierarchical structure of their elements, there are two
qualitatively different types of eigenstate that can arise. In the first case,
a neutrino can dominantly combine with its own antineutrino to form a
Majorana particle. The second case occurs when a neutrino combines
dominantly with an antineutrino, which is not the CP conjugate state, to form a
2-component massive neutrino. For example the electron neutrino might combine
with the muon antineutrino. Such states naturally occur in pairs with
order of magnitude-wise degenerate masses. In the example given, the other
member of the pair of Majorana states would be formed by combining the electron
neutrino with the muon antineutrino. The hierarchical structure
which gives rise to this second case is of course ruled out phenomenologically
for the quark and charged lepton mass matrices, as none have a pair of states
with order of magnitude-wise degenerate masses. However, considering
two generations for simplicity, a neutrino mass matrix of the form
\begin{equation}
M_{\nu} = \bordermatrix{	 &  \nu_1	&  \nu_2	\cr
			\overline{\nu}_1 &     0	&	  B		\cr
			\overline{\nu}_2 &	   B	&	  A		\cr}
\end{equation}
with the assumed hierarchy
\begin{equation}
|B| \gg |A|
\end{equation}
could be phenomenologically relevant. The mass eigenvalues are
$m_1 = B + A/2$ and $m_2 = B - A/2$, giving a neutrino mass squared difference
$\Delta m^2 = 2AB$, and the neutrino mixing angle is
$\theta \simeq \pi/4$ giving maximal mixing.
Maximal neutrino mixing, $\sin^2 2\theta \simeq 1$, provides a candidate
explanation \cite{halzen,smirnov} for
(i) the atmospheric muon neutrino deficit with
$\Delta m^2 = 10^{-2} eV^2$ and $\nu_{\mu}$-$\nu_{\tau}$ oscillations, or
(ii) the solar neutrino problem with $\Delta m^2 = 10^{-10} eV^2$ and
$\nu_{e}$-$\nu_{\mu}$ vacuum oscillations.

\section{Chiral Flavour Symmetries and Mass Protection}
It is natural to try to explain the occurrence of large mass ratios in terms of
selection rules due to approximate conservation laws.
A Dirac mass term:
\begin{equation}
-m \overline{\psi}_R \psi_L + h.c.
\end{equation}
connects a left-handed fermion component $\psi_L$ to its right-handed partner
$\psi_R$. If $\psi_L$ and $\psi_R$ have different quantum numbers, i.e. belong
to
inequivalent irreducible representations (IRs) of a symmetry group $G$ ($G$ is
then called a {\em chiral\/} symmetry), then the mass term is forbidden in the
limit of exact $G$ symmetry and they represent two massless Weyl particles. $G$
thus ``protects'' the fermion from gaining a mass.
Note that this is exactly the situation for all the SM
fermions, which are mass-protected by $SU(2)_L
\times U(1)_Y$ (but not by $SU(3)_c$). The $SU(2)_L
\times U(1)_Y$ symmetry is spontaneously broken and the  SM fermions gain
masses
suppressed relative to the presumed fundamental (GUT or Planck) mass scale $M$
by
the symmetry breaking parameter:
\begin{equation}
\epsilon = \langle\phi_{WS}\rangle/M
\end{equation}
The extreme smallness of this parameter $\epsilon$ constitutes, of course, the
gauge hierarchy problem.

Here we are interested in the further suppression of the quark and lepton mass
matrix elements relative to $\langle\phi_{WS}\rangle$.
We take the view \cite{fn1} that this hierarchy is due to the existence
of further approximately conserved chiral quantum numbers beyond those of the
SMG. The SMG is then a
low energy remnant of some larger group $G$ and the fermion mass and mixing
hierarchies are consequences of the spontaneous breaking of $G$ to the SMG.
The mass matrix element
suppression factors depend on how the fermions behave {\em w.r.t.\/} $G$
and on the symmetry breaking mechanism itself.

Consider, for example, an $SMG \times U(1)_f$ model, whose fundamental mass
scale
is M, broken to the SMG by the VEV of a scalar field $\phi_S$ where
$\langle\phi_S\rangle <  M$ and $\phi_S$ carries  $U(1)_f$ charge $Q_f(\phi_S)$
=
1.  Suppose further that $Q_f(\phi_{WS})=0$, $Q_f(b_L)=0$ and $Q_f(b_R)=2$.
Then
it is natural to expect the generation of a $b$ mass of order:
\begin{equation}
\left( \frac{\langle\phi_S\rangle }{M} \right)^2\langle\phi_{WS}\rangle
\end{equation}
via (see Fig.~\ref{figure}) the exchange of two $\langle\phi_S\rangle$
tadpoles,
\begin{figure}[hbt]
\begin{center}
\setlength{\unitlength}{1mm}
\begin{picture}(80,40)(0,15)
\put(0,45){\line(1,0){80}}
\put(20,45){\line(0,-1){20}}
\put(40,45){\line(0,-1){19}}
\put(60,45){\line(0,-1){20}}
\put(18.3,24){$\times$}
\put(38.3,24){$\otimes$}
\put(58.3,24){$\times$}
\put(17,17){$\langle\phi_{S}\rangle$}
\put(36,17){$\langle\phi_{WS}\rangle$}
\put(57,17){$\langle\phi_{S}\rangle$}
\put(4,47){$Q_f = 0$}
\put(24,47){$Q_f = 1$}
\put(44,47){$Q_f = 1$}
\put(64,47){$Q_f = 2$}
\put(0,40){$\large{b}_L$}
\put(76,40){$\large{b}_R$}
\put(29,40){M}
\put(49,40){M}
\end{picture}
\end{center}
\par
\fcaption{Feynman diagram which generates the b quark mass via
superheavy intermediate states.}
\label{figure}
\end{figure}
in addition to the usual
$\langle\phi_{WS}\rangle$ tadpole, through two appropriately charged
vector-like
superheavy (i.e. of mass M) fermion intermediate states \cite{fn1}.
We identify
\begin{equation}
\epsilon_f=\frac{\langle\phi_S\rangle }{M}
\end{equation}
as the $U(1)_f$ flavour symmetry breaking parameter.
 In general we expect mass matrix elements of order
\begin{equation}
M(i,j)\simeq \epsilon_{f}^{n_{ij}}\langle\phi_{WS}\rangle
\label{eq:mij}
\end{equation}
where
\begin{equation}
 n_{ij}= \mid Q_f(\psi_{L_{i}}) - Q_f(\psi_{R_{j}})\mid
\label{eq:nij}
\end{equation}
is the degree of forbiddenness due to the $U(1)_f$ quantum number difference
between the left- and right-handed fermion components. So the {\em effective\/}
SM Yukawa couplings of the quarks and leptons to the Weinberg-Salam Higgs
\begin{equation}
y_{ij}\simeq \epsilon_{f}^{n_{ij}}
\label{eq:yij}
\end{equation}
can consequently be small even though all {\em fundamental\/} Yukawa couplings
of
the ``true'' underlying theory are of $\cal O$(1).
We are implicitly  assuming here that
there exists a superheavy spectrum of states which can mediate all of the
symmetry breaking transitions; in particular we
do not postulate the {\em absence\/} of appropriate
superheavy states in order to obtain exact texture zeroes in the mass matrices
\cite{dim}.
We now consider models based on this idea.

Recently a systematic analysis of fermion masses in SO(10) SUSY-GUT
models has been made \cite{anderson} in terms of effective
operators obtained by integrating out the superheavy states, which
are presumed to belong to vector-like SO(10) {\bf16} + {$\bf \overline{16}$}
representations, in tree diagrams like Fig.~\ref{figure}. The minimal
number of effective operators contributing to mass matrices consistent
with the low energy data is four, which leads to the consideration of
GUT scale Yukawa coupling matrices satisfying Yukawa
unification, Eq.~(\ref{yukun}), and having the following texture
\begin{equation}
Y_i = \pmatrix{0               & z_{i}'C   		 & 0\cr
		      z_{i}C  		  & y_{i}Ee^{i\phi}  & x_{i}'B\cr
		      0               & x_{i}B           & A\cr}
\end{equation}
where i = u, d, l. Here the $x_{i}$, $x_{i}'$, $y_{i}$, $z_{i}$
and $z_{i}'$ are SO(10) Clebsch Gordon coefficients. These Clebschs
can take on a very large number of discrete values, which are
determined once the set of 4 effective operators (tree diagrams)
is specified. A scan of millions of operators leads to just 9
solutions consistent with experiment, having Yukawa coupling matrices
with a partial Georgi-Jarlskog structure of the form:

\begin{equation}
Y_u =\pmatrix{0               & \frac{-1}{27}C   & 0\cr
		      \frac{-1}{27}C  & 0                & x_{u}'B\cr
		      0               & x_{u}B           & A\cr}
\,
Y_d =\pmatrix{0				  & C				 & 0\cr
			  C				  & Ee^{i\phi}		 & x_{d}'B\cr
			  0				  & x_{d}B			 & A\cr}
\,
Y_l =\pmatrix{0				  & C				 & 0\cr
			  C				  & 3Ee^{i\phi}		 & x_{l}'B\cr
			  0				  & x_{l}B			 & A\cr}
\end{equation}
 For each of the 9 models
the Clebschs $x_{i}$ and $x_{i}'$ have fixed values and
the Yukawa matrices depend on 6 free parameters: A, B, C, E,
$\phi$ and $\tan\beta$. Each solution gives 8 predictions consistent
with the data, as illustrated in Table \ref{table:hall} for one of the
models.

\vspace{-1pt}
\begin{table}[h,t,b]
\centering
\tcaption{ Predictions for Model 6
with $\alpha_s(M_Z) = 0.115$. The so-called
Bag constant $\hat{B_K} $ has been determined
by lattice calculations to be in the range
$\hat{B_K} = 0.7 \pm 0.2$.}
\vspace{0.2cm}
\begin{tabular}{|c|c|c|c|}
\hline
  Input Quantity & Input Value & Predicted Quantity & Predicted Value\cr
\hline
  $m_b(m_b)$ & $4.35 $ GeV & $M_t$ & $176$ GeV \cr
  $m_\tau(m_\tau)$ &$1.777 $ GeV & $\tan\beta$ & $55 $  \cr
\hline
  $m_c(m_c)$ &$1.22$ GeV & $V_{cb}$ & $.048 $ \cr
\hline
  $m_\mu$ &$105.6 $ MeV   & $V_{ub}/V_{cb}$ & $.059 $ \cr
  $m_e $   &$0.511$ MeV  &  $m_s(1GeV)$ & $172 $ MeV \cr
  $V_{us}$       &$0.221 $    & $\hat{B_K} $ & $0.64$  \cr
                 &            & $m_u / m_d$  & $0.64$  \cr
                 &            & $m_s / m_d$  & $24.$  \cr
\hline
\end{tabular}
\label{table:hall}
\end{table}

The parameter hierarchy $A \gg B$, $E \gg C$ and the texture zeros are
interpreted as due to an approximately conserved global $U(1)_f$ symmetry
and the chosen superheavy fermion spectrum.  The global $U(1)_f$
charges are assigned in such a way that only the 4 selected
tree diagrams are allowed. In particular the
texture zeros reflect the assumed absence of
superheavy fermion states which could mediate the transition between the
corresponding Weyl states. A more detailed analysis of this $U(1)_f$
flavour symmetry is promised \cite{anderson}.

We now turn to models in which the chiral flavour charges are part of the
extended gauge group. The values of the chiral charges are
then strongly constrained by the anomaly conditions for the gauge theory.
It will also be assumed that any superheavy state needed to mediate a symmetry
breaking transition exists, so that the results
are insensitive to the details of the superheavy spectrum.
Consequently there will be no exact texture zeros but just highly suppressed
elements given by expressions like Eq.~(\ref{eq:mij}). The aim in these
models is to reproduce all quark-lepton masses and mixing angles within
a factor of 2 or 3.

The $SMG \times U(1)_f$ model obtained by extending the SM with a gauged
abelian flavour group appears \cite{bijnens} unable to explain the
fermion masses and mixings using an anomaly-free set of flavour charges.
Models extending the SM (or the MSSM) with discrete gauge symmetries
and having new interactions at energies as low as 1 Tev have also been
investigated \cite{leurer}

In a recent paper \cite{ibanezross}, Ibanez and Ross consider the extension
of the MSSM by an abelian flavour group $U(1)_f$. They then consider the
construction of an anomaly free \mbox{$MSSM \times U(1)_f$} model
having quark mass matrices with a texture
very close to that of solution 2 in Table~\ref{table:ross}. The
quarks and leptons are assigned the following $U(1)_f$ charges:
\begin{equation}
\left( \begin{array}{ccccc}
d_L \;&\; u_R \;&\; d_R \;&\; e_L \;&\; e_R \\
s_L \;&\; c_R \;&\; s_R \;&\; \mu_L \;&\; \mu_R \\
b_L \;&\; t_R \;&\; b_R \;&\; \tau_L \;&\; \tau_R
       \end{array} \right)
=\left( \begin{array}{ccccc}
-4 \;&\;  4 \;&\;  4 \;&\; -7/2 \;&\;  7/2 \\
 1 \;&\; -1 \;&\; -1 \;&\;  1/2 \;&\; -1/2 \\
 0 \;&\;  0 \;&\;  0 \;&\;   0  \;&\;  0
       \end{array} \right)
\label{eq:ross}
\end{equation}
Since the charge assignments are axial, the quark
and charged lepton mass matrices are symmetric
up to factors of order unity.
In addition to the
two Higgs doublets of MSSM, which are taken to be neutral under $U(1)_f$, two
Higgs singlets, $\theta$ and $\bar{\theta}$, are introduced with $U(1)_f$
charges $+1$ and $-1$ respectively and equal vacuum expectation values.
The $U(1)_f^2 U(1)_Y$ gauge anomaly vanishes.
The $U(1)_f^3$ anomaly  and the mixed $U(1)_f$ gravitational anomaly could
be cancelled against spectator particles neutral under the SMG.
However cancellation
of the mixed $SU(3)^2 U(1)_f$, $SU(2)^2 U(1)_f$ and $U(1)_Y^2 U(1)_f$
anomalies is only possible in the
context of superstring theories via the Green Schwarz mechanism
\cite{greenschwarz} with $sin^2\theta_W = 3/8$.
Consequently the $U(1)_f$ symmetry is spontaneously broken slightly
below the string scale.

 The $U(1)_f$ charge assignments of Eq.~\ref{eq:ross}
generate Yukawa matrices , via Eq.~\ref{eq:yij}, of the following form:
\begin{equation}
Y_u \simeq \pmatrix{\epsilon^8  	 & \epsilon^3    	& \epsilon^4\cr
		      		\epsilon^3  	 & \epsilon^2   	& \epsilon\cr
		      		\epsilon^4  	 & \epsilon   		& 1\cr}
\quad
Y_d \simeq \pmatrix{\bar{\epsilon}^8 & \bar{\epsilon}^3	& \bar{\epsilon}^4\cr
		      		\bar{\epsilon}^3 & \bar{\epsilon}^2 & \bar{\epsilon}\cr
		      		\bar{\epsilon}^4 & \bar{\epsilon}	& 1\cr}
\quad
Y_l \simeq \pmatrix{\bar{\epsilon}^5 & \bar{\epsilon}^3 & 0\cr
		      		\bar{\epsilon}^3 & \bar{\epsilon}  	& 0\cr
		      		0  				 & 0   				& 1\cr}
\end{equation}
The correct order of magnitude for all the masses and mixing angles are
obtained by fitting $\epsilon$, $\bar{\epsilon}$ and $\tan\beta$. This
is a large $\tan\beta \approx m_t/m_b$ model, but not necessarily
having exact Yukawa unification.

In the antigrand unified model \cite{holger,book}, the fundamental non-simple
gauge group $SMG^3 \equiv SMG_1 \times SMG_2 \times SMG_3$ (where
each factor $SMG_a$ acts non-trivially only on the a'th generation)
breaks down near the Planck scale to the usual SMG. This model has
several broken chiral flavour charges, corresponding to the gauge
subgroups $SU_a(3), SU_a(2)$ and $U_a(1)$, which can suppress fermion
mass matrix elements \cite{book}.
Any matrix element affected by a particular approximately
conserved non-abelian subgroup will be suppressed by the same factor,
because all suppressed transitions are identical
({\em triplet} $\leftrightarrow$ {\em singlet} for $SU_a(3)$ or
{\em doublet} $\leftrightarrow$ {\em singlet} for $SU_a(2)$). However
the matrix elements affected by an abelian subgroup $U_a(1)$ are
not suppressed identically, since the differences in the a'th generation
weak hypercharge between the corresponding left- and right-handed Weyl
components vary. The overall suppression of the mass matrix elements can
be written in the form:
\begin{equation}
M(i,j) \quad = \quad y_{ij}^{non-ab} \: y_{ij}^{ab}
\: \langle \phi_{WS} \rangle
\label{eq:gerryansatz}
\end{equation}
The non-abelian contributions are given by:
\begin{equation}
Y_u^{non-ab} \simeq Y_d^{non-ab} \simeq
\pmatrix{\epsilon_1 & \epsilon_2\delta_1\delta_2 &
\epsilon_3\delta_1\delta_3\cr
		 \epsilon_1\delta_1\delta_2 & \epsilon_2 & \epsilon_3\delta_2\delta_3\cr
		 \epsilon_1\delta_1\delta_3 & \epsilon_2\delta_2\delta_3 & \epsilon_3\cr}
\quad
Y_l^{non-ab} \simeq \pmatrix{\epsilon_1 & \epsilon_2 & \epsilon_3\cr
		      				 \epsilon_1 & \epsilon_2 & \epsilon_3\cr
							 \epsilon_1 & \epsilon_2 & \epsilon_3\cr}
\end{equation}
where $\epsilon_a$ is the symmetry breaking parameter for $SU_a(2)$ and
$\delta_a$ is the symmetry breaking parameter for $SU_a(3)$.
A natural measure of the degree of suppression by the abelian $U_a(1)$
components is given by the distance in abelian charge space
parameterised by a general metric $g_{ab}$:
\begin{equation}
y_{ij}^{ab} \quad = \quad
\exp [-\sqrt{(Q_{ai} - Q_{aj}) g_{ab} (Q_{bi} - Q_{bj})} \:]
\end{equation}
where $Q_{ai}$ is the value of the a'th generation weak hypercharge
carried by the i'th Weyl state.

The above ansatz Eq.~(\ref{eq:gerryansatz}) can readily explain the
generation mass gaps but not the mass splittings within each generation,
as it inevitably predicts \cite{fln:np1}:
\begin{equation}
m_u m_c m_t \: \simeq \:m_e m_{\mu} m_{\tau} \: \leq \: m_d m_s m_b
\end{equation}
So we are led to extend the gauge group further and $SMG^3 \times U(1)_f$ is
the only non-trivial anomaly-free extension with no new fermions and
the $U(1)_f$ charges are essentially unique:

\begin{equation}
\left( \begin{array}{ccccc}
d_L \;&\; u_R \;&\; d_R \;&\; e_L \;&\; e_R \\
s_L \;&\; c_R \;&\; s_R \;&\; \mu_L \;&\; \mu_R \\
b_L \;&\; t_R \;&\; b_R \;&\; \tau_L \;&\; \tau_R
       \end{array} \right)
=\left( \begin{array}{ccccc}
0 \;&\;  0 \;&\;  0 \;&\; 0 \;&\;  0 \\
0 \;&\;  1 \;&\; -1 \;&\; 0 \;&\; -1 \\
0 \;&\; -1 \;&\;  1 \;&\; 0 \;&\;  1
       \end{array} \right)
\label{eq:u1soln}
\end{equation}

\begin{table}[h,b,t]
\centering
\tcaption{Results of an $SMG^3 \times U(1)_f$ model fit to
		 fermion masses and mixing angles. All masses are
         running masses evaluated at 1 GeV unless otherwise stated.
         The third column shows a fit biased in favour
         of obtaining $m_c > m_s$.}
\vspace{0.2cm}
\begin{tabular}{||c||c|c|c||}
\hline
Fit & $m_t^{phys}=100\;GeV$ & \multicolumn{2}{c||}
{$m_t^{phys}=200\;GeV$} \\
\cline{3-4}
Results & & unbiased & biased \\
\hline \hline
$\chi^2$ & 3.7 & 5.6 & 6.9 \\ \hline \hline
$m_e$ (MeV) & 1.0 & 1.0 & 1.0 \\ \hline
$m_{\mu}$ (MeV) & 120 & 160 & 110 \\ \hline
$m_{\tau}$ (GeV) & 1.4 & 1.5 &1.5 \\ \hline \hline
$m_d$ (MeV) & 4.9 & 4.9 & 4.9 \\ \hline
$m_s$ (MeV) & 600 & 790 & 530 \\ \hline
$m_b^{phys}$ (GeV) & 5.4 & 5.5 & 5.3 \\ \hline \hline
$m_u$ (MeV) & 4.9 & 4.9 & 4.9 \\ \hline
$m_c$ (GeV) & 0.73 & 0.53 & 0.84 \\ \hline \hline
$V_{us}$ & 0.19 & 0.22 & 0.22 \\ \hline
$V_{cb}$ & 0.016 & 0.012 & 0.0048 \\ \hline
$V_{ub}$ & 0.0030 & 0.0027 & 0.0027 \\
\hline
\end{tabular}
\label{tab:dndiag}
\end{table}

A good order of magnitude fit to the data can
now be obtained \cite{fln:np1}
using 5 degrees of freedom and results are shown in the first column of
Table \ref{tab:dndiag}. All the data are fitted
within a factor of 2, except for $m_s$ and $V_{cb}$ which are fitted within a
factor of 3.

\section{Conclusion}
All the fermions except the top quark are light compared to the electroweak
scale $\langle \phi_{WS} \rangle$. So we might obtain a dynamical
understanding of $m_t$ - the SUSY fixed point value is particularly
promising - before understanding the electron mass and the rest
of the spectrum. The large top to bottom quark mass ratio is a mystery,
which can be exchanged for the mystery of a hierarchy of Higgs vacuum
values; all the third generation masses are then consistent with
quasifixed point values and/or Yukawa unification. However, in this
large $\tan\beta$ scenario, SUSY radiative corrections to $m_b$ are
generically large. There exist several mass matrix ans\"{a}tze with
texture zeros giving typically 5 successful relations between mass and
mixing parameters (including the 3 Georgi Jarlskog GUT scale relations:
$m_b = m_{\tau}$, $m_s = m_{\mu}/3$ and $m_d = 3m_e$). However these
ans\"{a}tze merely incorporate the mass hierarchy. Their hierarchical
structure strongly suggests the existence of approximately conserved
chiral gauge quantum numbers beyond those of the SMG responsible
for mass protection.

\section{References}

\end{document}